\documentclass[12pt,a4paper]{article}

\usepackage[utf8]{inputenc}
\usepackage{geometry}
\usepackage{authblk} 
\usepackage{amssymb}
\usepackage{amsmath,bbm}
\usepackage{psfrag}
\usepackage{graphicx}
\usepackage[verbose,hypertexnames=false]{hyperref}

\newcommand{\Z}[1]{\ensuremath{\mathbbm{Z}_{#1}}} 

\title{New features in the Z2xZ2 3HDM two component DM model}

\author{
  Jorge C. Romão\thanks{jorge.romao@tecnico.ulisboa.pt}, 
  Rafael Boto\thanks{rafael.boto@tecnico.ulisboa.pt},
  Pedro N. Figueiredo\thanks{pedro.m.figueiredo@tecnico.ulisboa.pt},
  João P. Silva\thanks{jpsilva@cftp.tecnico.ulisboa.p} 
}

\affil{Instituto Superior Técnico, Departamento de Física and CFTP\\
  A. Rovisco Pais 1,  1049-001, Lisboa, Portugal
}

\begin{document}

\maketitle

\begin{abstract}
We investigate\footnote{This talk presented at the conference
    \textit{My Favourite Dark Matter Model} Apr 14-17, 2025, Azores,
    summarizes the work
    published in Ref.\cite{Boto2024}.}
 the constraints and phenomenology of a three Higgs
doublet model (3HDM) with a $\Z2\times\Z2$ symmetry, featuring two
inert scalar doublets that give rise to a two-component dark matter
(DM) scenario. We analyze the model's vacuum structure, exploring the
competition between different symmetry-breaking minima, and subject it
to comprehensive theoretical and current experimental constraints. Our
analysis reveals previously unexplored regions of parameter space with
viable dark matter candidates. Notably, we identify scenarios where
both DM particles contribute comparably to the observed relic density,
offering distinctive experimental signatures that could guide future
searches.
\end{abstract}


\section{Introduction}

N-Higgs Doublet Models (NHDM) offer simple yet effective extensions of
the Standard Model (SM) with a variety of new features. These
models allow for the possibility of CP violation, provide viable
candidates for Dark Matter (DM), and possess large regions of
parameter space that are experimentally accessible, particularly at
the Large Hadron Collider (LHC).

A common variant is the 2HDM, which introduces besides the SM a new
scalar doublet with a $\Z2$ symmetry. This symmetry leaves the SM
fields unchanged and leads to the Inert Doublet Model (IDM) scenario. In
this setup, the DM particle states were initially in thermal
equilibrium with the primordial plasma before decoupling via the
freeze-out mechanism. The relic abundance of DM, given by
$\Omega_{\text{DM}}h^2 \approx 0.12$ (Planck, 2021)\cite{Planck},
is determined through the thermally averaged cross-section $\langle
\sigma v \rangle$. The viable DM mass range in this model typically
lies between $M_{h}/2 \lesssim M_{\rm DM} \lesssim M_{h}$, or
alternatively, $500\, \text{GeV} \lesssim M_{\rm DM}$.

In a more complex extension, the Two Inert 3HDM introduces a $\Z2
\times \Z2$ symmetry, which again does not alter the SM fields but
enforces additional constraints. This symmetry forbids the decay of
one sector to the other, further enhancing the stability of the DM
candidate and providing new avenues for study in both direct and
indirect detection experiments. 

We discuss in detail all possible vacua, and the conditions guaranteeing that
the double inert vacuum is indeed the global minimum. We also discuss all
the  collider, astrophysical and cosmological constraints which we
will use in our simulations. We show that it is possible to have a DM
candidate mass at any value $[\tfrac{1}{2}m_h, 1000 \text{GeV}]$, while it is
possible that both candidates contribute equally to the relic density
for either
$\tfrac{1}{2}m_h <  m_{H_1} < 80\, \text{GeV}$ or $m_{H_1}\gtrapprox 500\,
\text{GeV}$. 

\section{The Model}

We consider the $\Z2\times\Z2$ 3HDM. The potential is written
as\cite{Boto:2022uwv},

\begin{minipage}{1.0\linewidth}
\begin{align}
V =& m_{11}^2\phi_1^\dagger\phi_1 + m_{22}^2\phi_2^\dagger\phi_2 + m_{33}^2\phi_3^\dagger\phi_3+ \lambda_1(\phi_1^\dagger\phi_1)^2
+\lambda_2(\phi_2^\dagger\phi_2)^2+\lambda_3(\phi_3^\dagger\phi_3)^2\nonumber\\
&+
\lambda_4(\phi_1^\dagger\phi_1)(\phi_2^\dagger\phi_2)
+\lambda_5(\phi_1^\dagger\phi_1)(\phi_3^\dagger\phi_3)+\lambda_6(\phi_2^\dagger\phi_2)(\phi_3^\dagger\phi_3)
+\lambda_7(\phi_1^\dagger\phi_2)(\phi_2^\dagger\phi_1)
\nonumber\\
& +\lambda_8(\phi_1^\dagger\phi_3)(\phi_3^\dagger\phi_1)
+\lambda_9(\phi_2^\dagger\phi_3)(\phi_3^\dagger\phi_2)
\nonumber\\
&
+\left[\lambda''_{10}(\phi_1^\dagger\phi_2)^2 +
  \lambda''_{11}(\phi_1^\dagger\phi_3)^2 +
  \lambda''_{12}(\phi_2^\dagger\phi_3)^2
  +
\text{h.c.}\right]\, .
\end{align}
\end{minipage}

\section{Model Consistency}

\subsection{BFB}

The first thing we have to check is if the potential is bounded from
below(BFB). To this end, and also to look at the possible minima we
use the parameterization of Ref.\cite{Faro:2019vcd},
    \begin{equation}
    \label{eq:1}
    \phi_1=\sqrt{r_1}
    \begin{pmatrix}
        \sin\alpha_1\\
        \cos\alpha_1\ e^{i \beta_1}
    \end{pmatrix},\quad
    \phi_2= \sqrt{r_2}
    e^{i \gamma}
    \begin{pmatrix}
        \sin\alpha_2\\
        \cos\alpha_2\ e^{i \beta_2}
    \end{pmatrix},\quad
    \phi_3=\sqrt{r_3}
    \begin{pmatrix}
        0\\
        1
    \end{pmatrix} .
    \end{equation}
The necessary and sufficient conditions for the
$\Z2\times\Z2$ 3HDM to be bounded from below are only known
along neutral directions \cite{Grzadkowski:2009bt},
$\alpha_1=\alpha_2=0$.  
Only sufficient when considering charge breaking directions\cite{Faro:2019vcd}.
We have derived\cite{Boto:2022uwv} a general method of obtaining sufficient
conditions for 3HDM's. To obtain these we start by writing the 
quartic part of the potential as
\begin{equation}
   V_4=V_N + V_{CB} + V_{\Z2\times\Z2} ,
\end{equation}
where
\begin{align}
V_N =&  \lambda_1(\phi_1^\dagger\phi_1)^2
+\lambda_2(\phi_2^\dagger\phi_2)^2+\lambda_3(\phi_3^\dagger\phi_3)^2
+
(\lambda_4+\lambda_7)(\phi_1^\dagger\phi_1)(\phi_2^\dagger\phi_2)\nonumber\\
&
+(\lambda_5+\lambda_8)(\phi_1^\dagger\phi_1)(\phi_3^\dagger\phi_3)
+(\lambda_6+\lambda_9)(\phi_2^\dagger\phi_2)(\phi_3^\dagger\phi_3)\, ,
\\
V_{CB}=& -\lambda_7 z_{12}
-\lambda_8 z_{13}
-\lambda_9 z_{23} \, ,
\\
V_{\Z2\times\Z2}=&
\left[\lambda''_{10}(\phi_1^\dagger\phi_2)^2 +
  \lambda''_{11}(\phi_1^\dagger\phi_3)^2 +
  \lambda''_{12}(\phi_2^\dagger\phi_3)^2
  +
\text{h.c.}\right]\, ,
\end{align}
where
\begin{equation}
  \label{eq:3}
0\leq z_{ij}= 
  (\phi_i^\dagger\phi_i) (\phi_j^\dagger\phi_j)
  - (\phi_i^\dagger\phi_j) (\phi_j^\dagger\phi_i)\leq r_i r_j \, .
\end{equation}
Now the idea is to find a potential lower than the original one and
for which it is simple to find the BFB conditions. These will be
sufficient but not necessary conditions.
\begin{align}
 &V_4 \geq  V_{4}^{\rm lower}= V_N + V_{CB}^{\rm lower} + V_{\Z2\times\Z2}^{\rm lower}
  \equiv \frac{1}{2} \sum_{ij} r_i A_{ij} r_j\, ,\\
 &V_{CB}^{\rm lower} = r_1 r_2\ \text{min}(0,-\lambda_{7})
+ r_1 r_3\ \text{min}(0,-\lambda_{8})
+ r_2 r_3\ \text{min}(0,-\lambda_{9})\, , \\[+2mm]
&  V_{\Z2\times\Z2}^{\rm lower} =- 2|\lambda''_{10}| r_1 r_2
- |\lambda''_{11}| r_1 r_3
- |\lambda''_{12}| r_2 r_3\, ,
\end{align}
where the $A$ matrix is,
\begin{equation}
  A=\hskip -2mm
  \begin{bmatrix}
    \bar{\lambda}_{11}& \bar{\lambda}_{12}& \bar{\lambda}_{13}\\
    \bar{\lambda}_{12}& \bar{\lambda}_{22}&\bar{\lambda}_{23}\\
    \bar{\lambda}_{13}& \bar{\lambda}_{23}& \bar{\lambda}_{33}\\
  \end{bmatrix}\, ,
\end{equation}
and
\begin{align}
  & \bar{\lambda}_{11}= 2\lambda_1,\ \bar{\lambda}_{22}=
  2\lambda_2,\ \bar{\lambda}_{33}= 2\lambda_3\, ,\\
  &\bar{\lambda}_{12}=\lambda_4 + \lambda_7 +
  \text{min}(0,-\lambda_{7}) -  2|\lambda''_{10}| \, , \\
  &\bar{\lambda}_{13}=\lambda_5 + \lambda_8 +
   \text{min}(0,-\lambda_{8}) -  2|\lambda''_{11}| \, ,\\
   &\bar{\lambda}_{23}=\lambda_6 + \lambda_9 +
   \text{min}(0,-\lambda_{9}) -  2|\lambda''_{12}|\, .
 \end{align}
The potential is BFB if the matrix $A$ is copositive. The copositivity
conditions for a $3\times 3$ matrix
can be found in Ref.\cite{Kannike:2012pe}.

\subsection{Vacua}

We are interested in a \texttt{2-Inert} minimum, $(0,0,v)$. We need
to identify the parameter space with the \texttt{2-Inert}
configuration as the global minimum, having for all other minima, 
\begin{equation}
  \label{eq:2}
  V_{\texttt{2Inert}} < V_{\texttt{X}}\, .\nonumber   
\end{equation}

We now list all the possible minima, either among the neutral
directions or along the charge breaking ones.

\subsubsection{Possible Vacua: The Neutral case }

We start with the vacua along the neutral directions. The most general
neutral vacuum configuration may be parameterized as, 

\begin{equation}
  \langle \phi_1 \rangle  =
  \begin{bmatrix}
    0\\
    v_1 e^{i\xi_1}
  \end{bmatrix},\quad
  \langle \phi_2 \rangle = 
  \begin{bmatrix}
    0\\
    v_2 e^{i\xi_2}
  \end{bmatrix},\quad
  \langle \phi_3 \rangle =
  \begin{bmatrix}
    0\\
    v_3
  \end{bmatrix} .
\end{equation}
This was studied in Ref.\cite{Hernandez-Sanchez:2020aop}. We use
their notation 
for the naming of the different vacua, as shown in Tab.~\ref{tab:1}.

\begin{table}[htb]
  \begin{tabular}{c}
\scalebox{1}{
\begin{minipage}{0.75\linewidth}
  \begin{tabular}{|c|c|c|c|}\hline
    Name &vevs &Symmetry& Properties\\\hline
    \texttt{EWs}&(0,0,0)&$\Z2\times\Z2^\prime$ &EW Symmetry\\\hline
  \texttt{2-Inert}&$(0,0,v_3)$&$\Z2\times\Z2^\prime$ &SM + 2 DM candidates\\\hline
    \texttt{DM1}&$(0,v_2,v_3)$&$\Z2$ &2HDM + 1 DM candidates\\\hline
\texttt{DM2}&$(v_1,0,v_3)$&$\Z2^\prime$ &2HDM + 1 DM candidates\\\hline
\texttt{F0DM1}&$(0,v_2,0)$&$\Z2$ &1 DM candidates + massless fermions
    \\\hline
\texttt{F0DM2}&$(v_1,0,0)$&$\Z2^\prime$ &1 DM candidates + massless fermions
    \\\hline
\texttt{F0DM0}&$(v_1,v_2,0)$&None &No DM candidate + massless fermions
    \\\hline
\texttt{N}&$(v_1,v_2,v_3)$&None &3HDM no DM candidate
    \\\hline
\texttt{sCPv}&$(v_1e^{i \xi_1},v_2e^{i \xi_2},v_3)$&None &Spontaneous
CP violation\\\hline
\multicolumn{4}{c}{ }\\[-2mm]\hline
\texttt{F0DM0'}&$(v_1,i v_2,0)$&None &No DM candidate + massless fermions
\\\hline
\end{tabular}
\end{minipage}
}    
  \end{tabular}
  \caption{List of the neutral vacua.}
  \label{tab:1}
\end{table}

\vspace{3mm}

We have confirmed their vacua but we have found a new one
that we called $\texttt{F0DM0'}$. 

\subsubsection{Possible Vacua: The Charged case}

Now we turn to the charge breaking vacua. We
continue to follow the notation of
Ref.\cite{Hernandez-Sanchez:2020aop}. The list of all possible
charge breaking vacua are shown
in Tab.~\ref{tab:2}.
\begin{table}[htb]
  \centering
  \begin{tabular}[r]{r}
\scalebox{0.9}{\hskip 30mm
\begin{minipage}{0.6\linewidth}
\begin{tabular}{|c|c|}\hline
  Name &vevs \\ \hline
\texttt{CB1}&
$
\left( \begin{array}{c}
u_1\\
c_1
\end{array} \right)
\ \ 
\left( \begin{array}{c}
u_2\\
c_2
\end{array} \right)
\ \ 
\left( \begin{array}{c}
0\\
c_3
\end{array} \right)
$
 \\ \hline
  \texttt{CB2}&
$
\left( \begin{array}{c}
u_1\\
0
\end{array} \right)
\ \ 
\left( \begin{array}{c}
u_2\\
c_2
\end{array} \right)
\ \ 
\left( \begin{array}{c}
0\\
c_3
\end{array} \right)
$
 \\\hline
    \texttt{CB3}&
$
\left( \begin{array}{c}
u_1\\
c_1
\end{array} \right)
\ \ 
\left( \begin{array}{c}
u_2\\
0
\end{array} \right)
\ \ 
\left( \begin{array}{c}
0\\
c_3
\end{array} \right)
$
\\\hline
\texttt{CB4}&
$
\left( \begin{array}{c}
u_1\\
c_1
\end{array} \right)
\ \ 
\left( \begin{array}{c}
u_2\\
c_2
\end{array} \right)
\ \ 
\left( \begin{array}{c}
0\\
0
\end{array} \right)
$
 \\\hline
\texttt{CB5}&
$
\left( \begin{array}{c}
0\\
c_1
\end{array} \right)
\ \ 
\left( \begin{array}{c}
u_2\\
c_2
\end{array} \right)
\ \ 
\left( \begin{array}{c}
0\\
c_3
\end{array} \right)
$
 \\\hline
\texttt{CB6}&
$
\left( \begin{array}{c}
u_1\\
c_1
\end{array} \right)
\ \ 
\left( \begin{array}{c}
0\\
c_2
\end{array} \right)
\ \ 
\left( \begin{array}{c}
0\\
c_3
\end{array} \right)
$
 
    \\\hline
\texttt{CB7}&
$
\left( \begin{array}{c}
u_1\\
0
\end{array} \right)
\ \ 
\left( \begin{array}{c}
u_2\\
0
\end{array} \right)
\ \ 
\left( \begin{array}{c}
0\\
c_3
\end{array} \right)
$

    \\\hline
\texttt{CB8}&
$
\left( \begin{array}{c}
u_1\\
0
\end{array} \right)
\ \ 
\left( \begin{array}{c}
0\\
0
\end{array} \right)
\ \ 
\left( \begin{array}{c}
0\\
c_3
\end{array} \right)
$

    \\\hline
\texttt{CB9}&
$
\left( \begin{array}{c}
0\\
0
\end{array} \right)
\ \ 
\left( \begin{array}{c}
u_2\\
0
\end{array} \right)
\ \ 
\left( \begin{array}{c}
0\\
c_3
\end{array} \right)
$
 \\\hline
\multicolumn{2}{c}{ }\\\hline
\texttt{F0CB}&
$
\left( \begin{array}{c}
u_1\\
c_1
\end{array} \right)
\ \ 
\left( \begin{array}{c}
u_2\\
-\frac{u_1^* u_2}{c_1^*}
\end{array} \right)
\ \ 
\left( \begin{array}{c}
0\\
0
\end{array} \right)
$

\\\hline
  \end{tabular}
\end{minipage}
}
\end{tabular}
  \caption{List of the charge breaking vacua.}
  \label{tab:2}
\end{table}
As before, we have identified an extra possibility; the last entry in
the table.  To confirm our results we have also numerically minimized
the potential with \texttt{Minuit}\cite{james:1975dr} for random
initial conditions to make sure all possibilities were verified.

\subsubsection{Global minima procedure:}

To find the correct global minimum we therefore follow the steps,

\begin{itemize}
\item Identify all possible minima.
\item Apply BFB and conditions for \texttt{2-Inert} to be global minima.
\item  Numerically minimize the potential with
  \texttt{Minuit} for random initial conditions.
\item Confirm none of the points ever give a lower vacuum.
\item In particular we verified that the sufficient BFB were correct
  has we did not find any point unbounded from below.
\end{itemize}

\subsubsection{Other Constraints}

The remaining restrictions to consider include,
\begin{itemize}
\item The S matrix must satisfy perturbative
  unitarity (use the results from\cite{Bento:2022vsb} for all the 3HDMs).

\item Agreement with the S, T and U electroweak precision
  parameters using the formulas from\cite{Grimus:2008nb}.

\item 
  For $h_{125}$, we use
  $\mu_{if}^h$ from ATLAS\cite{ATLAS:2022vkf}
  \begin{figure}[htb]
    \centering
    \scalebox{0.75}{\hskip -45mm
      \begin{minipage}{1.0\linewidth}
        \includegraphics[scale=1]{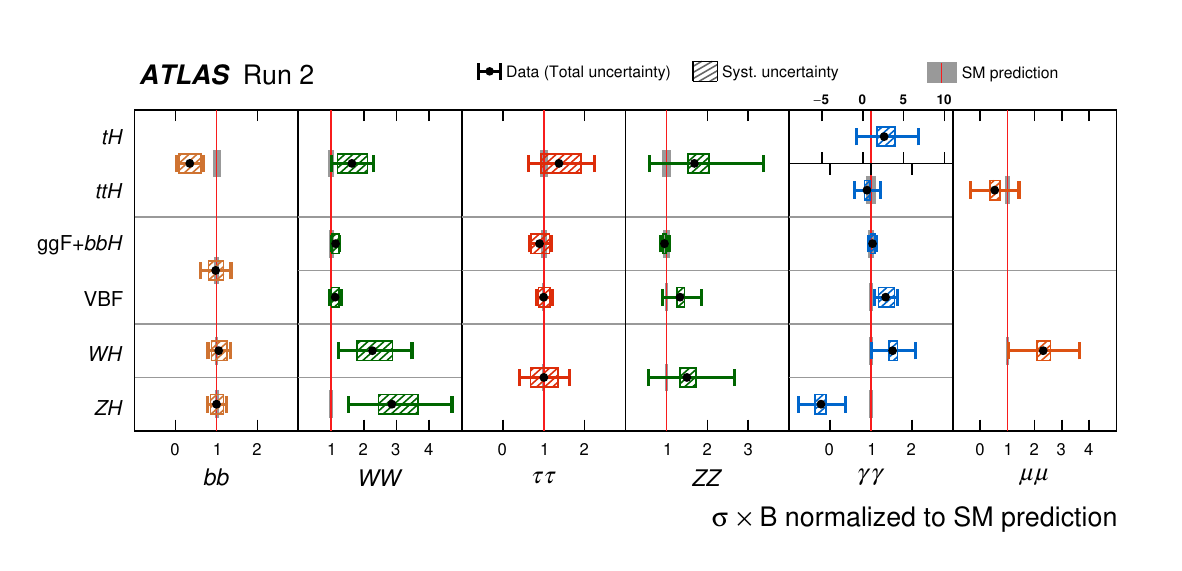}   
      \end{minipage}
    }    
    \caption{Signal strengths from ATLAS\cite{ATLAS:2022vkf}.}
    \label{fig:atlas}
  \end{figure}
  .

\item For the other scalars we use \texttt{HiggsTools} 1.1.3
  \cite{Bahl:2022igd}   that
      uses the experimental cross section limits from the LEP, the
      Tevatron and the LHC (at 95\% C.L). 
\item Upper limit on the Higgs total decay width is set by
  \cite{CMS:2019ekd} at $\Gamma_{tot}\leq 9.1\, \text{MeV}\,.$
\end{itemize}

\subsubsection{Constraints from Dark Matter}

Finally we have to comply with the constraints coming from DM. These are,
\begin{itemize}
\item 
  The total relic density\cite{Planck} is given by the sum of the
  contributions from the DM candidates,  
 \begin{equation}
     \Omega_T h^2 = \Omega_{H_1} h^2+ \Omega_{H_2} h^2 = 0.1200 \pm
   0.0012 .
 \end{equation}
\item The invisible Higgs branching ratio\cite{ATLAS:2023tkt},
  $Br(h\rightarrow \mbox{invisible})<0.11$.

\item  From direct detection - Spin-independent (SI) scattering cross
      section from LZ\cite{LZ:2022lsv} with future  DARWIN/XLZD and
      PandaX-xT.    Rescale the calculated $\sigma$ by the relative
      relic density. 
  \item From indirect detection - detect gamma rays, cosmic rays or
      neutrinos from DM annihilation.
    For GeV scale, use Fermi-LAT\cite{Fermi-LAT:2015att} $\gamma's$ in dwarf
    galaxies, antiproton from flux AMS\cite{AMS:2016oqu}  and
    $\gamma's$ in Galactic Center from H.E.S.S.\cite{HESS:2022ygk}. 
\end{itemize}

\section{The Scan}

To perform the scan it proves useful to define\cite{Hernandez-Sanchez:2020aop},
\begin{equation}
\Lambda_1 =
\tfrac{1}{2} \left( \lambda_4 + \lambda_7 + 2 \lambda''_{10} \right)\, ,
\Lambda_2 =
\tfrac{1}{2} \left( \lambda_6 + \lambda_9 + 2 \lambda''_{12} \right)\, ,
\Lambda_3 =
\tfrac{1}{2} \left( \lambda_5 + \lambda_8 + 2 \lambda''_{11} \right)  .
\end{equation}
The parameter space has 15 parameters that we choose as follows:
\begin{equation}\label{parameterspace}
    v^2, m_{H_1}^2, m_{H_2}^2, m_{H_3}^2=m_{SM}^2, m_{A_1}^2,
    m_{A_2}^2, m_{H_1^\pm}^2, m_{H_2^\pm}^2, \Lambda_1, \Lambda_2,
    \Lambda_3, \lambda_1, \lambda_2, \lambda_4, \lambda_7 .
\end{equation}
We choose random values in the ranges
\begin{align}
&\Lambda_1,\, \Lambda_2,\, \Lambda_3,\, \lambda_1,\, \lambda_2,\,
\lambda_4,\, \lambda_7  \in \pm\left[10^{-3},10\right]; 
\nonumber\\[2pt] 
&m_{H_1},\, m_{H_2},\, m_{A_1},\,m_{A_2}\,
\in \left[50,1000\right]\,\text{GeV};
\nonumber\\[2pt]
&
m_{H_1^\pm},\,m_{H_2^\pm}\,
\in \left[70,1000\right]\,\text{GeV},
\end{align}
We built a \texttt{FORTRAN} program for the model to calculate all the
quantities for a randomly generated set of parameters and test all the
constraints. We then generate the FeynRules and CalcHEP model files in
order to implement the model in micrOMEGAs 6.0.5. The couplings were
derived with the help of
\texttt{FeynMaster}\cite{Fontes:2019wqh,Fontes:2021iue,Fontes:2025svw}.

\section{Results}

\subsection{Direct detection}
\begin{figure}[htbp!]
\centering
\includegraphics[width=0.37\textwidth]{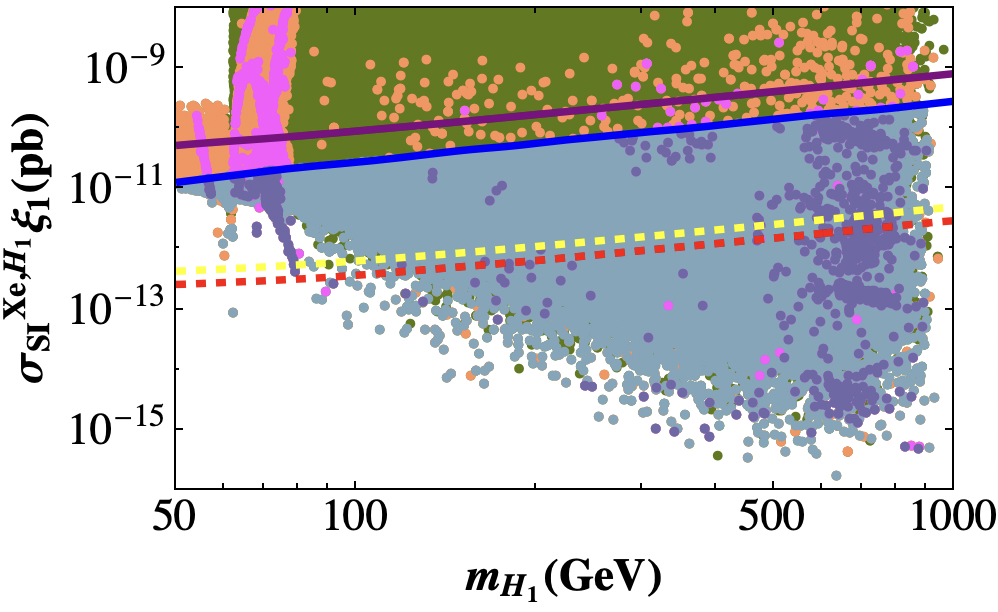}
\hspace{3mm}
\includegraphics[width=0.14\textwidth]{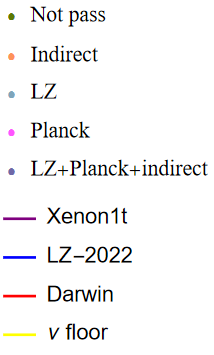}
\hspace{3mm}
\includegraphics[width=0.37\textwidth]{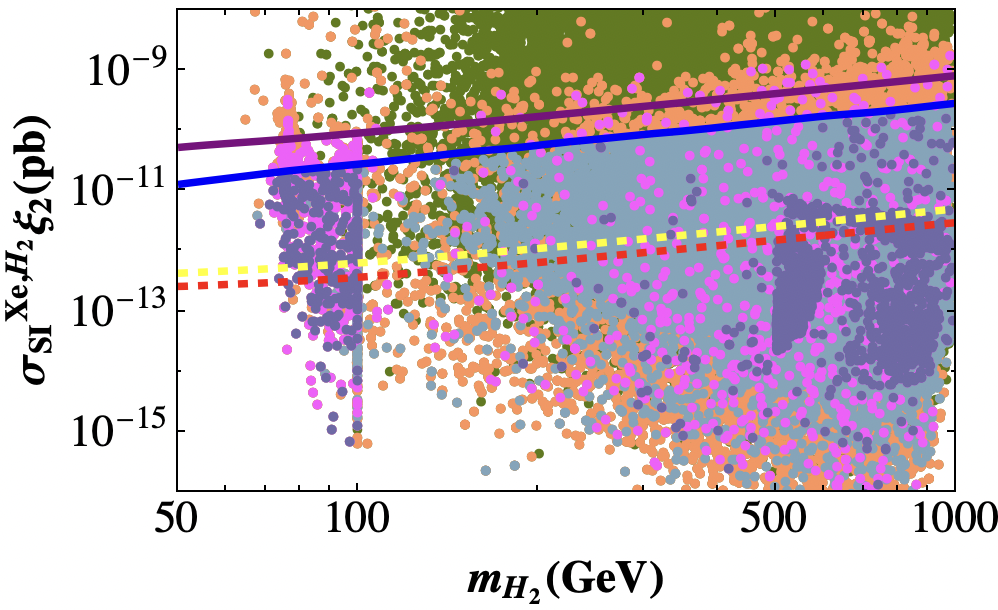}
\caption{Direct detection constraints on $H_1$ ($H_2$) on the left
  (right) figure. See text for details. 
}
\label{direct}
\end{figure}

In this section we present the results coming from applying all the
constraints. First we focus in direct detection. The results are shown
in Fig.~\ref{direct}.

The points pass all previous constraints, including collider and we
take $m_{H_2}>m_{H_1}$ always.  The presence of orange/pink points
above LZ line shows that this is a relevant exclusion method.  For low
$m_{H_1}$ it is possible that direct detection probes $H_1$ without
affecting $H_2$.  We also note that final exposures of DARWIN may
reach the high mass section of the neutrino floor. Then other probes
must be used.

\subsection{Indirect detection}

\begin{figure}[htbp!]
\centering
\includegraphics[width=0.38\textwidth]{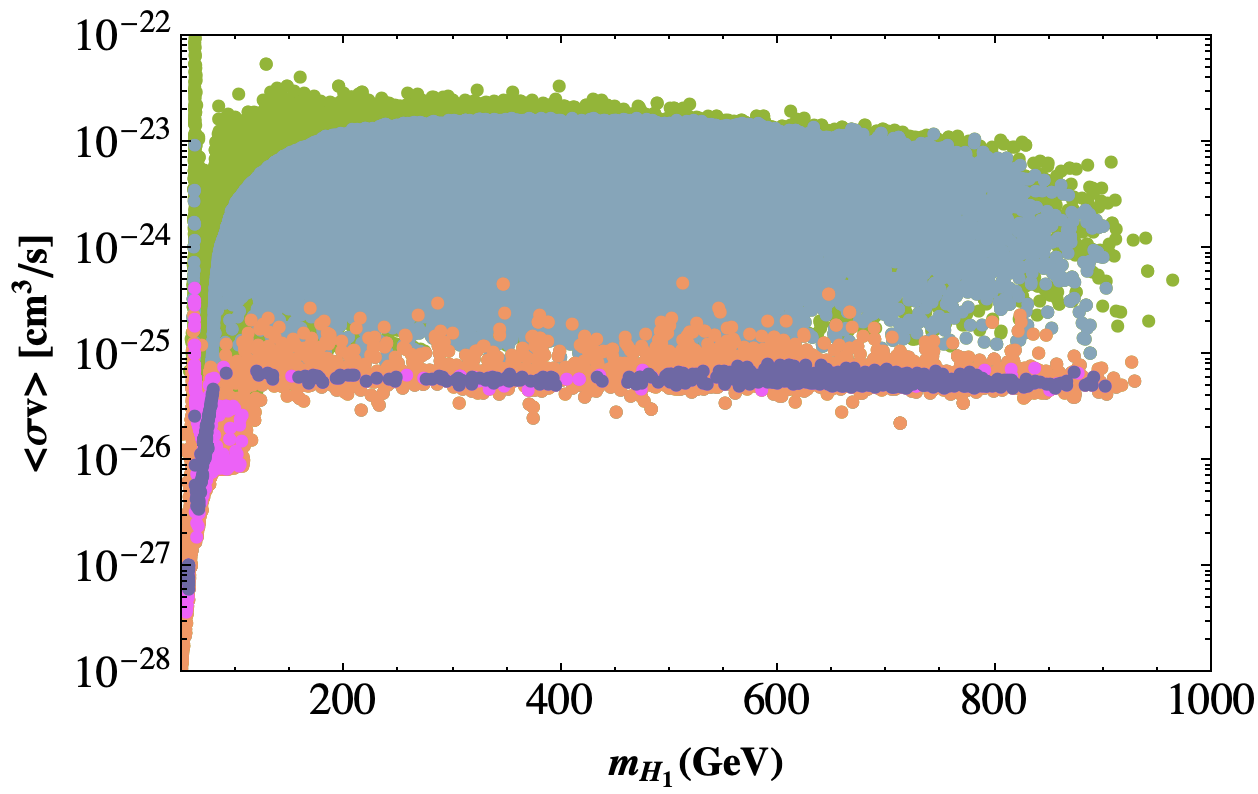}
\hspace{3mm}
\includegraphics[width=0.14\textwidth]{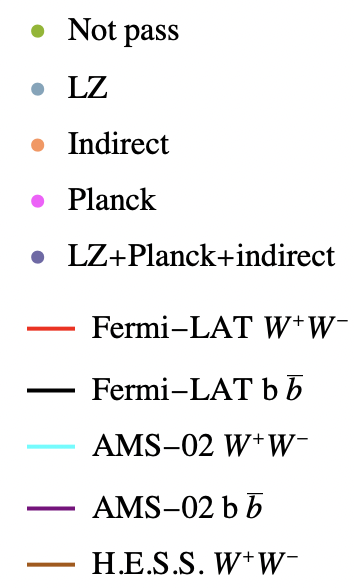}
\hspace{3mm}
\includegraphics[width=0.38\textwidth]{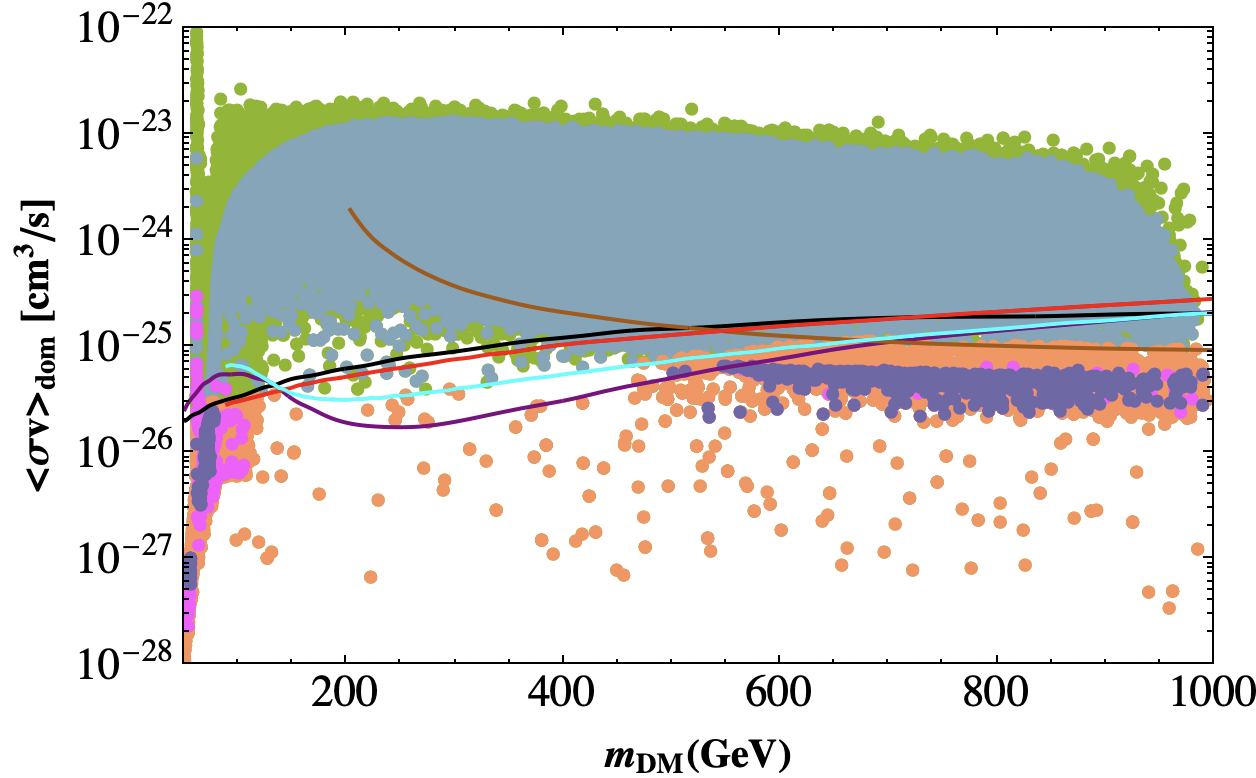}
\caption{The colours of the points have the same meaning as in Fig.~\ref{direct}.
The left figure shows the total
$\langle\sigma v\rangle$ as a function of $m_{H_1}$. 
The right figure shows the dominant contribution to
$\langle\sigma v\rangle$ as a function of the mass of the DM
candidate, $m_\textrm{DM}$, 
which corresponds to the $\langle\sigma v\rangle$ plotted on the vertical axis.
The lines coming from
Fermi-LAT \cite{Fermi-LAT:2015att} and H.E.S.S. \cite{HESS:2022ygk}
assume a Navarro-Frenk-White (NFW) 
DM density profile and the AMS-02 \cite{AMS:2016oqu} lines correspond
to the conservative approach derived in Ref.~\cite{Reinert:2017aga}, 
with the colour codes also shown in the figure. 
}
\label{indirect1}
\end{figure}

Now we turn to the indirect constraints.
Fig.~\ref{indirect1}-left shows the total $\langle\sigma v\rangle$.
The figure shows the total
$\langle\sigma v\rangle$ as a function of $m_{H_1}$. 
candidate. We assume a Navarro-Frenk-White (NFW) 
DM density profile.
We calculate the contributing channels and take the upper limit
from the reconstruction of the experimental signal for the dominant
channel:
above the $W$ threshold, the annihilation proceeds mostly into
$WW$; it occurs into $b \bar{b}$ otherwise. Importantly we have found
that the Planck constraints (almost) guarantee indirect detection to
not have effect.

\subsection{ Relic density}

\begin{figure}[htbp!]
\centering
\includegraphics[width=0.45\textwidth]{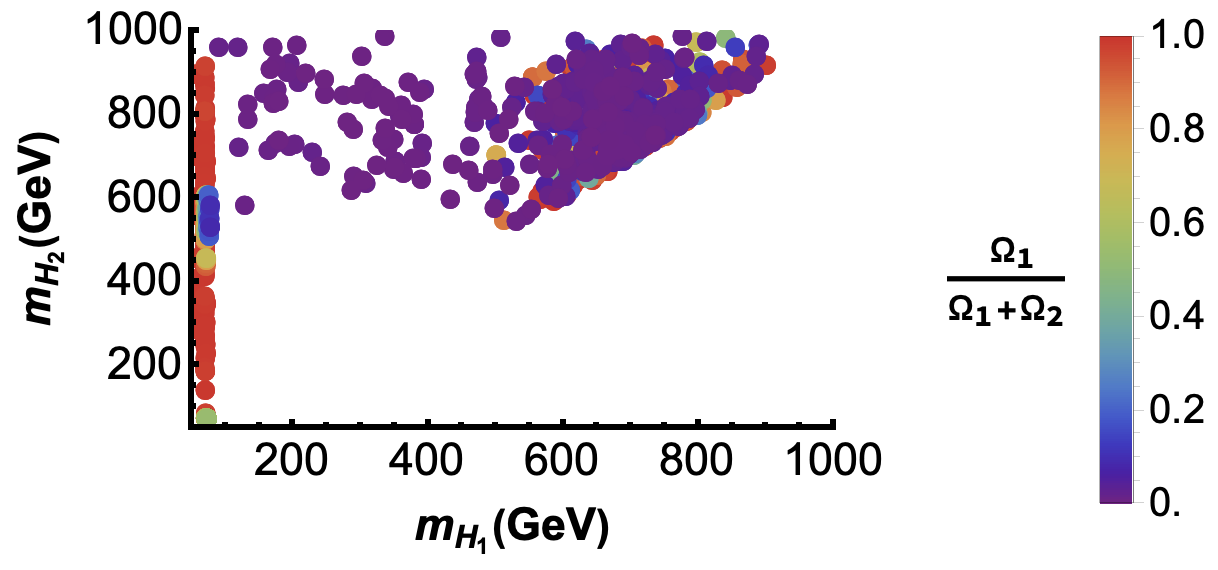}
\hspace{5mm}
\includegraphics[width=0.45\textwidth]{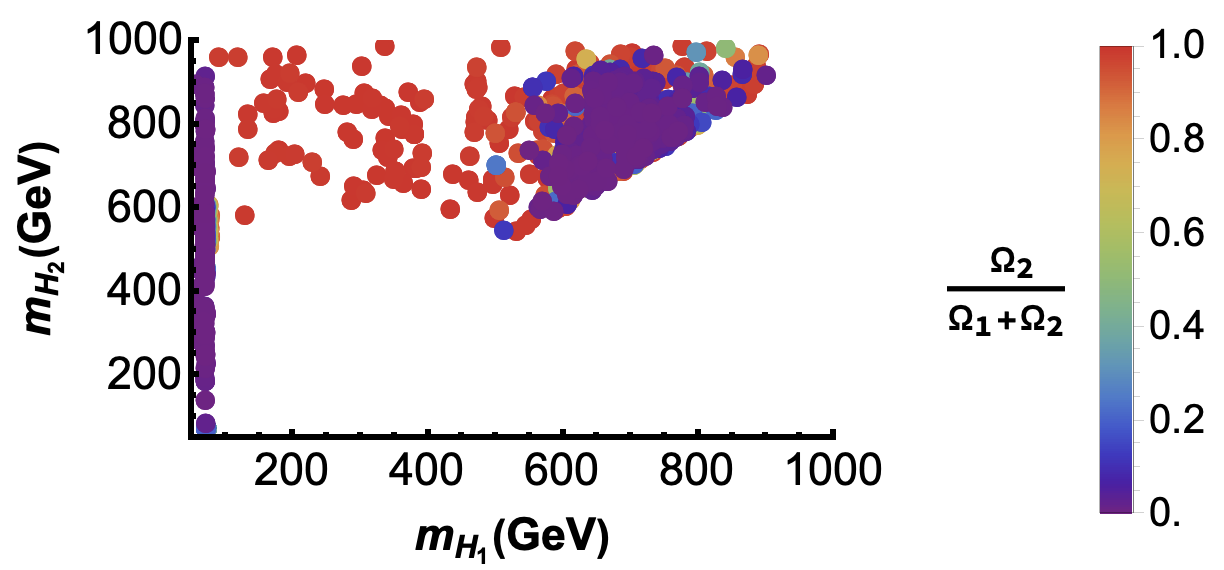}
\caption{Range of allowed ($m_{H_1}, m_{H_2}$) masses with a ``temperature''
colour code for $\Omega_1/\Omega_T$ ($\Omega_2/\Omega_T$) on the left (right).
}
\label{mH1mH2_relic}
\end{figure}

The results for the relic density are shown in Fig.~\ref{mH1mH2_relic}.
All points shown satisfy all constraints, including direct and indirect
detection, and have the correct total relic density.
If $H_1$ were the only DM component, combining relic density and
direct detection constraints would lead to the same mass regions as in
the IDM. It is possible to have a DM candidate mass at any value
$[\tfrac{1}{2}m_h, 1000 GeV]$ . In the intermediate mass range,
by requiring that it is $H_2$ which is mostly responsible for
the relic density. 
Equal abundance is possible for either $\tfrac{1}{2}m_h <
      m_{H_1} < 80 GeV$ or $m_{H_1}\gtrapprox 500 GeV$.  

\section{Conclusions}

In view of recent interest  in multi-component DM models,
we focus our attention on a  $\Z2\times\Z2$ symmetric 3HDM with a double
inert vacuum.
We begin by revisiting the potential solutions to the stationarity
equations, ensuring that our solution represents the absolute
minimum. In the process, we identify two new significant minima, which
we designate as \texttt{F0DM0}' and \texttt{F0CB}. 

To guarantee that we account for all global minima, we must compare
not only with the previously identified solutions \texttt{F0DM1},
\texttt{F0DM2}, and \texttt{F0DM0}, but also with these new minima:
\texttt{F0DM0}' and \texttt{F0CB}. 

We have derived explicit expressions for all these cases, which makes direct comparison straightforward. Additionally, we ensure consistency with key physical constraints, including unitarity, Bounded-From-Below (BFB) conditions, and compatibility with the oblique parameters 
$S,T$ and $U$.

Following this step, we apply all current collider constraints to the parameter space of our model, including limits on the 125 GeV couplings, searches for additional scalars, and flavor observables. Next, we focus on the implications of relic density, direct detection (DD), and indirect detection (ID) of dark matter (DM).

Through an extensive scan of the parameter space, we observe that the simplistic conclusions drawn from narrow regions of the parameter space no longer hold. Instead, a much broader and more diverse range of possibilities emerges. Notably, we find regions where two distinct DM candidates contribute equally to the relic density. In the $\Z2 \times \Z2$ model, the entire mass spectrum for a given component can be populated, including intermediate mass ranges where one component dominates the relic density calculation while the other plays a secondary role.

Additionally, we take into account the future sensitivity of direct detection experiments, which are expected to probe the high-mass section of the neutrino "fog" without ruling out the model. This highlights the need for complementary probes to fully test the viability of the model.

\section*{Acknowledgments}
This work is supported in part by the Portuguese FundaÃ§Ã£o para a
Ciencia e Tecnologia (FCT) through the PRR (Recovery and Resilience
Plan), within the scope of the investment ``RE-C06-i06 - Science Plus
Capacity Building'', measure ``RE-C06-i06.m02 - Reinforcement of
financing for International Partnerships in Science, Technology and
Innovation of the PRR'', under the project with reference
2024.01362.CERN.  The work is
also supported by FCT under Contracts UIDB/00777/2020, and
UIDP/00777/2020. The FCT projects are partially funded through POCTI
(FEDER), COMPETE, QREN, and the EU. The work of R. Boto is also
supported by FCT with the PhD grant PRT/BD/152268/2021.


\end{document}